\documentclass[fleqn,useAMS,usenatbib]{mnras}
\usepackage{fullpage}
\usepackage{bm}
\usepackage{amsmath}
\usepackage{amssymb}
\usepackage{hyperref}
\usepackage{graphicx}

\def\d{{\rm d}}
\def\e{{\rm e}}
\def\i{{\rm i}}

{\newif\ifnotend
\notendtrue
\def\veclist{ABCDEFGHIJKLMNOPQRSTUVWXYZabcdfghijklmnopqrstuvwxyz.}
\def\top#1#2.{#1}
\def\tail#1#2.{#2.}
\loop\expandafter\xdef\csname v\expandafter\top\veclist\endcsname%
{{\noexpand\bf\expandafter\top\veclist}}
\edef\veclist{\expandafter\tail\veclist}
\if\veclist.\notendfalse\fi\ifnotend\repeat}

\let\boldgrk=\gkvecten
\let\boldgrksc=\gkvecseven

\def\gkthing#1{{\mathchoice%
    {\hbox{{\boldgrk\char#1}}}
    {\hbox{{\boldgrk\char#1}}}
    {\hbox{{\boldgrksc\char#1}}}
    {\hbox{{\boldgrksc\char#1}}}}}

\def\vtheta{\gkthing{18}}

\def\vOmega{\gkthing{10}}

\def\stim{^{\rm e}}
\def\resp{^{f}}
\def\Omegap{\Omega_{\rm p}}
\def\massbar{m_{\rm b}}

\usepackage{graphicx}

\title{On the self-consistent time-dependent linearized response of
	stellar discs to external perturbations}
\author[Dominic Dootson]{Dominic Dootson and John Magorrian\\
	Rudolf Peierls Centre for Theoretical Physics, Clarendon
	Laboratory, Parks Road, Oxford OX1 3PU}

\begin{document}
	\maketitle
	\begin{abstract}
		We study the explicitly time-dependent response of a razor-thin axisymmetric disc to externally imposed perturbations
		by recasting the linearized Collisionless Boltzmann equation as an integral equation and applying Kalnajs' matrix method
                As an application we consider the idealized
                problem of calculating the dynamical friction torque
                on a steadily rotating, two-dimensional bar.
 		We consider two choices of basis functions in the matrix
                method, showing that both lead to comparable results.
                The torques from our linearised calculation are in
                excellent agreement with those measured from $N$-body
                simulation, as long as the bar perturbation does not resonate with a significant fraction of the disc's stars.
	\end{abstract}
	

	\section{Introduction}
	\label{Sec: Introduction}

        $N$-body simulation is the single most powerful technique for probing collisionless stellar dynamics.
        Although $N$-body simulations are superficially easy to run, setting them up and interpreting their results is often be difficult: they rely on Monte Carlo sampling the galaxy's phase space distribution function \citep[e.g.,][]{leeuwin1993n}, which can require prohibitively large~$N$ to ensure that both it and the gravitational potential it sources are sufficiently well sampled.
        Nevertheless, much of our understanding of stellar dynamics comes from carefully controlled $N$-body simulations \cite[e.g.,][]{sellwood2012spiral,SellwoodSpiralinstabilitieslinear2021}.
        
        Another approach is to use perturbation theory, focusing -- at least initially -- only on the linearized response of galaxies to perturbations.  The motivation for this is that the linear response should be easier to understand than the fullly nonlinear response produced by $N$-body simulation and ought to help us interpret the latter.
        Techniques for calculating the linear response are much less well developed than those for evolving $N$-body simulations though.
        The evolution of a galaxy involves two coupled equations: Poisson's equation and the collisionless Boltzmann equation (CBE).  In an $N$-body simulation these are typically both expressed in terms of familiar $(\vx,\vv)$ coordinates, but when studying perturbations of equilibrium galaxies it more natural to  solve the CBE using angle--action coordinates $(\vtheta,\vJ)$.  One then has to deal with the mapping between these and the original $(\vx,\vv)$ coordinates to solve Poisson's equation.  The construction of $(\vtheta,\vJ)$ coordinates and the map between them and $(\vx,\vv)$  is the main difficulty encountered in perturbative calculations.  We do not offer any new solutions to it in this paper.

        We address a more easily rectified perceived difficulty: with a few exceptions, most work in this area has used a Fourier transform in time to solve for the evolution.  This frequency-based approach is appropriate if one wants to test the stability of an isolated galaxy or to work out the response to an eternal, perfectly periodic perturbation \citep[e.g.][]{pichon1997numerical,evans1998stability,de2019instabilities}, but for more general scenarios it is an unnecessary obfuscation and it makes more sense to solve for the time evolution directly \citep[e.g.][]{julian1966non,Seguin1994,murali1999transmission, rozier2022constraining}.
        We explain the difference between these two approaches in Section~\ref{Sec: Dynamical Evolution}.
        As a simple example in Section~\ref{Sec: Application to razor-thin axisymmetric discs} we show how to calculate the explicitly time-dependent response of a razor-thing disc model, comparing the performance of two distinct choices of potential--density pairs \citep{kalnajs1976dynamics}.
        Section~\ref{Sec: Greens Function} presents a calculation of  the Green's function of the system, the object at the heart of all initial value analyses.
        As a straightforward application in Section~\ref{Sec: Perturbations by an External Galaxy} we calculate the dynamical friction torque exerted by the disc on an externally imposed bar, comparing the results to those measured in $N$-body simulations.  Appendices give details of the $N$-body code we use for this comparison and of the potential--density pairs used for the linear-response calculation.

	\section{Dynamical Evolution}
	\label{Sec: Dynamical Evolution}
	Consider a galaxy that is in equilibrium with its stars orbiting in a steady potential $\Phi(\vx)$ and having phase space mass density function $F(\vx,\vv)$.
	This equilibrium is perturbed by an externally imposed potential, $\epsilon\psi\stim(\vx,t)$.
	In response, the stars' DF changes from $F(\vx,\vv)$ to $F(\vx,\vv)+\epsilon f(\vx,\vv,t)$, with the change $\epsilon f$ in the DF adding $\epsilon\psi\resp(\vx,t)$ to the gravitational potential of the system.
	The subsequent evolution of the galaxy is governed by the CBE,
	\begin{equation}
	\frac{\partial }{\partial t}(F+\epsilon f) + \left[F+\epsilon f, H\right] = 0,
	\label{Equ: CBE}
	\end{equation}
	with Hamiltonian $H=H_0+\epsilon\psi$, in which we assume that $H_0\equiv \frac12\vv^2+\Phi$ and the perturbation potential
	\begin{equation}
	\psi=\psi\stim+\psi\resp,
	\end{equation}
	is the sum of the externally imposed stimulus~$\psi\stim$ plus the galaxy's response~$\psi\resp$.
	The latter is related to $f$ through Poisson's equation,
	\begin{equation}
	\label{eq:poisson}
	\nabla^2\psi\resp=4\pi G\int\d\vv f.
	\end{equation}
	
	We assume that angle-action variables $(\vtheta,\vJ)$ exist for the unperturbed galaxy's Hamiltonian~$H_0$. It is natural then to express the response DF $f$ as a Fourier series in the angles,
	\begin{equation}
	f(\vtheta,\vJ)=\sum_\vm f_\vm(\vJ)\e^{\i\vm\cdot\vtheta},\quad
        \label{eq:Fouriersum}
	\end{equation}
	with analogous expansions for $\psi\stim$ and $\psi\resp$.
	Invoking Jeans' theorem we may assume that $F=F(\vJ)$ only \citep{binney2011galactic}.
	Substituting these expansions into~\eqref{Equ: CBE} and dropping non-linear terms, those of $\mathcal{O}(\epsilon^2)$, the linearised evolution equation for the response DF becomes
	\begin{equation}
	\frac{\partial f_\vm(\vJ,t)}{\partial t} + \i\vm\cdot\left[\vOmega(\vJ) f_\vm(\vJ,t)
	- \frac{\partial F(\vJ)}{\partial\vJ} \psi_\vm(\vJ,t) \right] = 0,
	\label{Equ: FT CBE first order}
	\end{equation}
	where we have defined $\vOmega(\vJ) \equiv \frac{\partial H_0}{\partial\vJ}$.
	
	One way to proceed from here would be to Laplace transform in time in order to remove the derivative in the first term, resulting in a simple algebraic equation for the (transformed) response. As noted in the introduction, this is how the majority of work attempting solve the linearised CBE proceeds. Instead we follow the approach taken in \cite{murali1999transmission} and note that the formal solution to~\eqref{Equ: FT CBE first order}, using an integrating factor, is simply
	\begin{equation}
	f_\vm(\vJ,t) = \i\vm\cdot \frac{\partial F}{\partial\vJ}\int_{-\infty}^tdt'\,\e^{-\i\vm\cdot\vOmega(\vJ)(t-t')}\psi_\vm(\vJ,t').
	\label{Equ: CBE with integrating factor}
	\end{equation}
	The complication is that the $\psi_\vm=\psi\stim_\vm+\psi\resp_\vm$ factor in the integrand depends on the $f_\vm$ through Poisson's equation~\eqref{eq:poisson}.
	Following \cite{kalnajs1976dynamics}, we deal with this by introducing a complete set of potential--density pairs $(\psi^{(p)}(\vx),\rho^{(p)}(\vx))$, the elements of each pair~$p$ being related by Poisson's equation
	\begin{equation}
	\nabla^2\psi^{(p)}=4\pi G\rho^{(p)}.
	\end{equation}
	For any choice of basis there is a matrix
	\begin{equation}
	\mathcal{E}^{pq} = -\int\d\vx\left[\psi^{(p)}(\vx)\right]^{*}\rho^{(q)}(\vx),
	\label{Equ: Potential-density pair def}
	\end{equation}
	that describes how the elements of basis pairs $p$ and~$q$ are related.
	We expand the external potential and the induced response 
        \footnote{For brevity we adopt the convention that repeated
          basis-function indices, such as $p$ here, are summed over.   In constrast,
          sums over the Fourier indices $\vm$ (e.g.,
          equation~\ref{eq:Fouriersum}) will always be explicit.}
	\begin{subequations}
		\label{eq:ABexpansion}
		\begin{equation}
		\psi\stim(\vx,t) = A_{p}(t)\psi^{(p)}(\vx),
		\end{equation}
		\begin{equation}
		\psi\resp(\vx,t) = B_{p}(t)\psi^{(p)}(\vx),
		\end{equation}
\end{subequations}
	with coefficients $A_{p}(t)$ and $B_{p}(t)$ respectively.
          In principle the number of terms in this sum is infinite, but in order to make computational progress these expansions will be limited to a finite number of terms. \par 
	Given $\psi\stim(\vx)$ the coefficients $A_p$ that minimise the mean-square error
	$\int\left|\psi\stim-A_p\psi^{(p)}\right|^2\,\d\vx$ can be found by solving
	the system of of equations
	\begin{equation}
	\label{eq:potproj}
	\mathcal{P}^{pq}A_{q} = \int\d\vx\,\left(\psi^{(p)}\right)^\star \psi\stim,
	\end{equation}
	in which the matrix is
	\begin{equation}
	\begin{split}
	&\mathcal{P}^{pq} = \int d\vx\,\left(\psi^{(p)}\right)^\star\psi^{(q)}.\\
	\end{split}
	\end{equation}   
        Thanks to the linear nature of Poisson's equation the density that sources the potential response is given by
	\begin{equation}
	\rho\resp(\vx,t)=B_p(t)\rho^{(p)}(\vx). \\
	\end{equation}
	The expansion coordinates, $A_p(t)$ and $B_p(t)$, completely describe the evolution of our stimulus and response potentials, respectively.  
	\par

	We can now combine our potential density pair expansion, along with \autoref{Equ: CBE with integrating factor}, to get the evolution of the density response,
	\begin{equation}
	\mathcal{E}^{pq}B_q(t)=\int_{-\infty}^t \d t'\, \mathcal{K}^{pq}(t-t')\left[A_q(t')+B_q(t')\right].
	\label{Equ: Linear Volterra Equation of the Second Kind}
	\end{equation}
	This is a linear Volterra equation, where the kernel, $\mathcal{K}^{pq}$, is defined as
	\begin{equation}
	\begin{split}
	& \mathcal{K}^{pq}(t-t') = -\i(2\pi )^{d}\times\\
	& \int \d^{d}\vJ\sum_\vm \vm\cdot\frac{\partial F(\vJ)}{\partial\vJ} \e^{-\i\vm\cdot\vOmega(\vJ)(t-t')}
	\left(\hat{\psi}^{(p)}_{\mathbf{m}}(\mathbf{J})\right)^{*}
	\hat{\psi}^{(q)}_{\mathbf{m}}(\mathbf{J}),
	\label{Equ: Kernel}
	\end{split}
	\end{equation}
	and $d$ is the dimensionality of configuration space.
	It is straightforward to solve Equation~\eqref{Equ: Linear Volterra Equation of the Second Kind} numerically for the unknown response coefficients $B_q(t)$ once the kernel~\eqref{Equ: Kernel} and the orthogonality matrix~\eqref{Equ: Potential-density pair def}, $\mathcal{E}$, have been constructed. By introducing potential-density pairs \citep[following][]{kalnajs1976dynamics} we have reduced our problem to one of matrix arithmetic. \par 
	There is however, no free lunch. By expanding in potential-density pairs, we have embedded Laplace's equation into our formalism. Although this has allowed us to trivially include self-gravity, it has come at the cost: from \autoref{Equ: Linear Volterra Equation of the Second Kind} we can only solve for the real space response rather than the full phase space response. In principle, however, this problem can be rectified as once the potential evolution is determined, \autoref{Equ: CBE with integrating factor} can be solved to obtain the full phase space evolution.  \par 
	The kernel given in in \autoref{Equ: Kernel} has an intimate relationship to the response matrix, $\hat{\mathcal{M}}^{pq}(\omega)$ \citep[e.g.][]{binney2011galactic}, as it represents how a disk will respond to an external perturbation and its own response. In fact, an expression for the response matrix can be obtained by taking the two sided Laplace transform of \autoref{Equ: Linear Volterra Equation of the Second Kind}, swapping the order of integration to obtain,
	\begin{equation}
	\mathcal{E}^{pq}\hat{B_{q}}(\omega) = \hat{\mathcal{M}}^{pq}(\omega)\left[\hat{A_{q}}(\omega)+\hat{B_{q}}(\omega)\right],
	\end{equation}
	where the response matrix
	\begin{equation}
	\begin{split}
	& \hat{\mathcal{M}}^{pq}(\omega) =i(2\pi )^{d}\times \\
	&  \int  \d^{d}\vJ\sum_\vm \vm\cdot\frac{\partial F(\vJ)}{\partial\vJ}	\frac{\left(\hat{\psi}^{(p)}_{\mathbf{m}}(\mathbf{J})\right)^{*}
		\hat{\psi}^{(q)}_{\mathbf{m}}(\mathbf{J})}{\omega - i\vm \cdot\vOmega(\vJ)},
	\end{split}
	\label{Equ: Response Matrix}
      \end{equation}
	is the Laplace transformed kernel. 
	Although this reduces \autoref{Equ: Linear Volterra Equation of the Second Kind} to a simple matrix equation it introduces a vanishing denominator that causes much of the difficulties in the modal analysis. \par
	\section{Application to razor-thin axisymmetric discs}
	\label{Sec: Application to razor-thin axisymmetric discs}
	\begin{figure}
		\includegraphics[width = 0.5\textwidth]{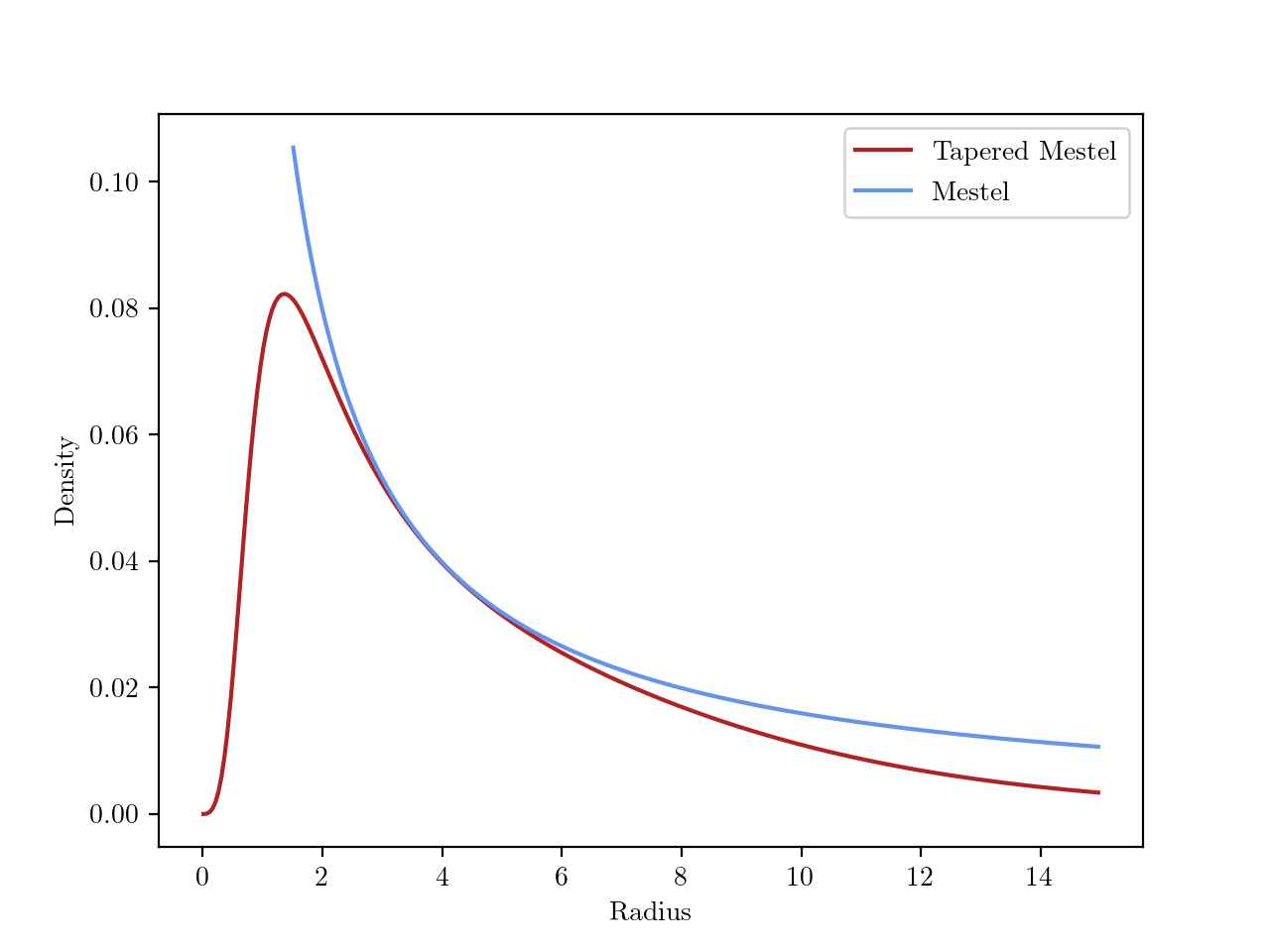}
		\caption{The density distribution for the tapered and untapered Mestel discs. The inner taper is at $R=1$ and the outer is at $R=10$.}
		\label{Fig: Background DF}
	\end{figure}
	In this paper we will restrict ourselves to in-plane perturbations of the razor-thin Mestel disc \citep{mestel1963galactic}.
	In equilibrium the disc's potential is $\Phi(R) = v_{c}^{2}\log (R/R_{0})$ where $v_{c}$ is the circular orbit speed and sets the mass scale of the disk.
	This potential is generated by a density profile $\rho_{\text{Mestel}}(R) = v_{c}^{2}/(2\pi G R)$, which is given by the DF, 
	\begin{equation}
	F_{\rm Mestel}(E,L) = \mathcal{N}L^{q}\e^{-E/\sigma_{R}^{2}},
	\label{Equ: Mestel DF}
	\end{equation}
	where $\sigma_{R}^{2}$ is proportional to the disc's temperature, $q \equiv v_{c}^{2}/\sigma_{R}^{2}-1$ and $\mathcal{N}$ is the normalisation,
	\begin{equation}
	\mathcal{N} = \frac{v_{c}^{2}}{2^{1+q/2}\pi^{3/2}G\sigma_{R}^{2}\Gamma(1/2+q/2)},
	\end{equation}
        chosen to make $\int d^{2}\mathbf{v}F_{\text{Mestel}}=\rho_{\text{Mestel}}$.
	In order to deal with the singularity at the centre of the disk and the infinite extent of the Mestel disk we follow the approach taken in \cite{sellwood2014transient} and \cite{fouvry2015secular} and introduce inner and outer tapers,
	\begin{equation}
	T_{\text{inner}}(L) = \frac{L^{\nu _{\rm t}}}{(R_{\rm i}v_{\rm c})^{\nu_{\rm t}}+L^{\nu_{\rm t}}},
	\label{Equ: Inner taper}
	\end{equation}
	\begin{equation}
	T_{\text{outer}}(L) = \left[1+ \left(\frac{L}{R_{\rm o}v_{\rm c}}\right)^{\mu_{\rm t}}\right]^{-1}.
	\label{Equ: Outer taper}
	\end{equation}
	The parameters $\nu_{\rm t}$ and $\mu_{\rm t}$ set the sharpness of the inner and outer taper respectively with $R_{\rm i}$ and $R_{\rm o}$ setting their location. We also assume that only a fraction $0<\xi\leq1$ of the disk is active.  This gives the total DF
	\begin{equation}
	F(E,L) = \xi T_{\text{inner}}(L)T_{\text{outer}}(L)F_{\rm Mestel}(E,L).
	\label{Equ: Distribution}
	\end{equation}
	We take the following parameters, $v_{\rm c} = G = R_{0} = 1$, this gives a natural timescale $t_{0} = R_{0}/v_{\rm c} =1$. For the distribution function we take the same parameters as \cite{sellwood2014transient} and \cite{fouvry2015secular}, namely: $\nu_{\rm t} = 4$, $\mu_{\rm t} = 5$, $R_{\rm i} = 1$, $R_{\rm o} = 10$ and $\xi = 1/2$. We take $\sigma_{r}=0.35$, unless otherwise specified.   \par
	Where appropriate we express times in terms of the dynamical time at the location of the inner taper -- the time taken to complete an circular orbit there.  This is $T_{\rm dyn} = 2\pi R_{\rm i}/v_{\rm c}=2\pi$. \par 

	\subsection{Potential basis functions}
	
	Taking advantage of the axisymmetry of the background DF we choose to decompose our basis functions as
	\begin{equation}
	\psi^{(p)}(r, \phi) =  \e^{\i\ell\phi}\mathcal{U}^{\ell}_{n}(R), \qquad \rho^{(p)}(R, \phi) = \e^{\i\ell\phi}\mathcal{D}^{\ell}_{n}(R),
	\label{Equ: Potential Basis}
	\end{equation}
	in which the label $p=(n,\ell)$.
	We assume that the  radial functions, $\mathcal{U}^{\ell}_{n}$ and $\mathcal{D}^{\ell}_{n}$, are real. If they are chosen well then we will be able to capture the relevant physics of the perturbation of interest using only a small number of coefficients, which will minimise the number of elements we need to include in the kernel $\mathcal{K}^{pq}$ (equation~\ref{Equ: Linear Volterra Equation of the Second Kind}). In this paper we make use of two different choices for these basis functions. \par 
		The first is the well-known \cite{kalnajs1976dynamics} basis, which is constructed to satisfy the biorthogonality relation 
	\begin{equation}
	-\frac{1}{2\pi}\int_0^{R_{\rm max}}\mathcal{D}_{n_1}^l(R)\mathcal{U}_{n_2}^l(R)\,R\d R=\delta_{n_1,n_2}.
	\end{equation}
        Then the matrix $\mathcal{E}^{pq}$ that appears on the right-hand side of~\eqref{Equ: Potential-density pair def} is simply $\mathcal{E}^{pq}=\delta^{pq}$.
        By construction, the density functions $\mathcal{D}_{n,l}(R)$ are equal to zero for radii $R>R_{\rm max}$.  We take $R_{\rm max}=20R_{0}$.  Appendix~\ref{App: Kalnajs Basis Functions} reviews this basis in more detail. 
	
	Our second choice is the Gaussian basis introduced by ~\cite{de2019instabilities}, for which the potential basis is given by
	\begin{equation}
	\mathcal{U}^\ell_{n}(R) =\frac{G\Delta_{n}^{2}}{R_{0}^{3}} \exp\left[-\frac{\left(R-R_{n}\right)^{2}}{2\Delta_{n}^{2}}\right],
	\label{Equ: Spiral Potential}
	\end{equation}
	independent of $\ell$, in which $R_n$ sets the centre of the $n^{\rm th}$ basis function and $\Delta_n$ its width.  The corresponding density is $\mathcal{D}_n^\ell(r)=-2\pi G\nabla^2\mathcal{U}_n^\ell(R)$.
Unlike the Kalnajs basis, these offer the possibility of tailoring the details of the basis to optimise the sampling both of the imposed perturbation and its response. We do not investigate that possibility here, but instead adopt a simpler approach. To ensure good sampling of the potential response at small radii, where the density is largest, we space $R_{n}$ logarithmically between $0.15R_{0}$ and $15R_{0}$.  The width, $\Delta_{n}$, is picked such that $R_{n}+\Delta_{n} = R_{n+1}$, making each basis function unresolved. Unlike the Kalnajs basis, these basis functions are not bi-orthonormal, but it is straightforward  to use Poisson's equation to find the density corresponding to each~\eqref{Equ: Spiral Potential} and then to use~\eqref{Equ: Potential-density pair def} to calculate $\mathcal{E}^{pq}$. \par
Given a basis, one must choose the order at which to curtail the expansion, $N_{max}$. The tapered Mestel disk's stability is highly sensitive to the inner taper \citep{evans1998stability}, the basis functions must therefore be able to resolve this fine structure at small radii; if not, instabilities can develop. \par 
One must also ensure that the perturbation can be sufficiently well resolved at larger radii. For the Kalnajs basis functions, for which the resolution increases linearly with $N_{max}$, ensuring good resolution of the inner taper will ensure a good resolution of any external perturbation over the whole disk. However for our choice of Gaussian basis functions, for which resolution increases only logarithmically with more basis functions, one has to pay more attention. By comparing the true perturbation with its representation in the basis functions one can ascertain if the number of basis functions needs to be increased. We discuss this matter further in section \ref{SubSec: Comparison of Different Basis Functions}, where we compare the two sets of basis functions.       


	%
        

	\subsection{Angle--action variables for 2d axisymmetric discs}
	\label{SubSec: 2D Basis Functions}
	
	After choosing a basis $\psi^{(p)}(\vx)$ the next step is to expand it as a Fourier series in angle--action coordinates,
	\begin{equation}
	\hat{\psi}^{(p)}_{\mathbf{m}}(\mathbf{J}) = \frac{1}{(2\pi)^{2}}\int \d\theta_{1} \d\theta_{2}\e^{-\i m_{1}\theta_{1}}
	\e^{-\i m_{2}\theta_{2}}\psi^{(p)}(R, \phi),
	\end{equation}
        in which $\vm=(m_1,m_2)$ and $R$ and $\phi$ vary as the angles $\theta_1,\theta_2$ advance along the orbit having actions~$\vJ$.
        Following \cite{fouvry2015secular} the actions are
	\begin{equation}
	J_{1} = J_{r} = \frac{1}{\pi}\int_{R_{-}}^{R_{+}}dr \sqrt{2(E - \Phi(R)) - L^{2}/R^{2}},
	\label{Equ: J_1 action}
	\end{equation}
	\begin{equation}
	J_{2} = J_{\phi} = L, 
	\label{Equ: J_2 action}
	\end{equation}
	where $(E,L)$ are the orbit's specific energy and angular momentum and $(R_{+},R_{-})$ are its apo- and peri-centre radii, respectively, which satisfy $2(E-\Phi(R))-\frac{L^{2}}{R^{2}}=0$.
	Each orbit has a pair of associated frequencies, $\Omega_{1}=\Omega_r$ and $\Omega_2=\Omega_{\phi}$, which are given by
	\begin{equation}
	\frac{2\pi}{\Omega_{1}} = 2\int_{R_{-}}^{R_{+}}\frac{dR}{\sqrt{2(E-\psi(R)) -L^{2}/R^{2}}},
	\end{equation}
	\begin{equation}
	\frac{\Omega_{2}}{\Omega_{1}} = \frac{L}{\pi}\int_{R_{-}}^{R_{+}}\frac{dR}{R^{2}\sqrt{2(E-\psi(R)) -L^{2}/R^{2}}}.
	\end{equation}
	Given $\psi(r)$ we can specify an orbit uniquely using any of $(R_{-},R_{+})$, $(E,L)$ or $(J_{r},J_{\phi})$.
	It is straightforward to map from one set of coordinates to the other, except for the map  $(J_{r},J_{\phi})\rightarrow (E,L)$ which requires the inversion of a non-linear integral map.
	In practice we transform the integral over $\vJ$ in equation~\eqref{Equ: Kernel} to one over $(R_-,R_+)$,  due to the trivial mapping to other coordinates and because the physical extent of the disk is more naturally expressed in $ (R_{-}, R_{+})$ than in $(E,L)$.
	Directly from the definition of apocentre and pericentre, we have that
	\begin{equation}
	E = \frac{R_{+}^{2}\Phi_{+} - R^{2}_{-}\Phi_{-}}{R^{2}_{+}-R^{2}_{-}}, \qquad L = \sqrt{\frac{2(\Phi_{+}-\Phi_{-})}{R_{-}^{-2} - R_{+}^{-2}}},
	\label{Equ: Apo & Peri centre}
	\end{equation}
	where $\Phi_{\pm} \equiv \Phi (R_{\pm})$.\par

	The angles associated with the actions \eqref{Equ: J_1 action} and~\eqref{Equ: J_2 action} are \citep{lynden1972generating}	
	\begin{subequations}
		\begin{equation}
		\theta_{1}(R,\phi) = \Omega_{1}\int_{C_{1}[R]}dR' \frac{1}{\sqrt{2(E-\Phi_{0}(R'))- L^{2}/R;'^{2})}},
		\end{equation}
		\begin{equation}
		\theta_{2}(R,\phi) = \phi + \int_{C_{1}[R]}dR' \frac{\Omega_{2} - L/R'^{2}}{\sqrt{2(E-\Phi_{0}(R')) - L^{2}/R'^{2}}},	
		\end{equation} 
	\end{subequations}
	in which the limits, $C_{1}$, goes from pericentre $R_{-}$ to the radius of interest, $R$.
	\cite{tremaine1984dynamical} showed that the basis functions~\eqref{Equ: Potential Basis} can be expressed as 
	\begin{equation}
	\hat{\psi}^{(n,\ell)}_{\mathbf{m}}(\mathbf{J}) 
	=
	\delta^{\ell}_{m_{2}}\mathcal{W}^{m_{1}}_{\ell m_{2} n}(\vJ),
	\label{Equ: Basis functions in action-angle space}
	\end{equation} 
	thanks to the symmetry assumed in our decomposition of the basis functions (\autoref{Equ: Potential Basis}). It is important to note that the Kronecker delta means that different angular harmonics decouple, a direct consequence of the decomposition in \autoref{Equ: Potential Basis}. If the radial function is real then $\mathcal{W}$ becomes,
	\begin{equation}
	\mathcal{W}^{m_{1}}_{\ell m_{2} n}(\vJ) = \frac{1}{\pi}\
	\int_{R_-}^{R_+}\d R\frac{\d\theta_{1}}{\d R}
	\mathcal{U}^{\ell}_{n}\cos\left[m_{1}\theta_{1} + m_{2}(\theta_{2}-\phi)\right],
	\label{Equ: Weyl Function}
	\end{equation}
	In transforming the~$\theta_1$ integral into one over~$R$ we have introduced a Jacobian
	\begin{equation}
	\frac{\d\theta_{1}}{\d R} = \frac{\Omega_{1}}{\sqrt{2(E-\Phi(R))- L^{2}/r^{2})}}. 
	\end{equation}
	The cosine in~\eqref{Equ: Weyl Function} is an even function, which means that $\mathcal{W}^{m_{1}}_{\ell m_{2}n} = \mathcal{W}^{-m_{1}}_{\ell -m_{2}n}$, reducing the number of calculations needed.
	We do the calculation of $\psi^{n\ell}_\vm$ for a  grid of points in $(R_-,R_+)$ spaced uniformly with spacing $\Delta R_{\text{grid}}$. The specific parameters required depend on the potential-density pairs used, we list these in detail in \autoref{App: Kalnajs Basis Functions} and \ref{Sec: Gaussian Basis Functions}. 
        
	\subsection{Computing the evolution Kernel}
	\label{SubSec: Evolution Kernel}
	Once we have the basis functions we must calculate the kernel for each time step from $t-t' = 0$ to $t-t' = t_{\rm end}$.
	As our basis functions are on a grid of points in $(R_{-}, R_{+})$ space we need to transform our integral measure and derivatives in~\eqref{Equ: Kernel} to reflect the change of coordinates.

	By the chain rule the derivative with respect to the background DF that appears in~\eqref{Equ: Kernel} is
	\begin{multline}
	\vm \cdot \frac{\partial F}{\partial\vJ} = m_{1}\Omega_{1}\left(\frac{\partial F}{\partial E}\right)_{L} \\+ m_{2}\left[\Omega_{2}\left(\frac{\partial F}{\partial E}\right)_{L} + \left(\frac{\partial{F}}{\partial L}\right)_{E}\right]. 
	\end{multline}
	We split the Jacobian associated with the change in integration variables from  $(J_{1}, J_{2}) \rightarrow (R_{-}, R_{+})$ into two factors.
	The first is the Jacobian from $(J_{1}, J_{2}) \rightarrow (E,L)$, which is given by
	\begin{equation}
	\frac{\partial(J_{1},J_{2})}{\partial(E,L)} = \frac{1}{\Omega_{1}}.
	\end{equation}
	The second is the Jacobian from $(E,L) \rightarrow (R_{-}, R_{+})$, which  is best calculated numerically from~\eqref{Equ: Apo & Peri centre}.
	The overall Jacobian is then the product of the two.\par
	
	Due to the time translation symmetry of the kernel we only calculate it at a set of points $t\in [0, \Delta t, 2\Delta t, ..., t_{\rm end}]$ where we use a uniform time step, $\Delta t$.
	We typically take $t_{\rm max}$ to be of order a few dynamical times.
	We can further reduced the number of calculations needed by exploiting the structure of equation~\eqref{Equ: Basis functions in action-angle space}.
	Due to the symmetries in the basis functions, the kernel satisfies the properties
	\begin{equation}
	\mathcal{K}_{n_{p}\ell_{q}}^{{n_{q}\ell_{q}}} (t-t') = \left[\mathcal{K}_{n_{p}\ell_{-q}}^{{n_{q}\ell_{-q}}} (t-t')\right]^{*},
	\end{equation}	
	\begin{equation}
	\mathcal{K}_{n_{p}\ell_{p}}^{{n_{q}\ell_{q}}} = \mathcal{K}_{n_{q}\ell_{q}}^{{n_{p}\ell_{p}}}.
	\end{equation}
	We can use these properties to reduce the time required to produce each kernel by approximately a factor of four.\par
	Once we have the kernel we can then use standard techniques to solve the integral equation~\eqref{Equ: Linear Volterra Equation of the Second Kind}.
	We use the trapezium rule to approximate the integral on the right-hand side by a sum.
	Rearranging this sum gives an explicit expression for the response coefficients $B_p(t_n)$ in terms of their earlier values: 
	\begin{equation}
	\begin{split}
	B_{s}(t_n) = &\Delta t\left[\mathcal{E}-\frac{1}{2}\mathcal{K}(t_{n})\right]^{-1}_{sp}\times\\
	&\bigg[\frac12\mathcal{K}^{pq}(t_n)\left(A_q(0) + B_q(0)\right)
	+ \\
	&\sum_{i=1}^{n-1}\mathcal{K}^{pq}(t_n-t_i)\left[A_q(t_i)+B_q(t_i)\right]
	\bigg].
	\end{split}
	\label{Equ: Numerical Volterra Equation}  \end{equation}
	%
	Here we have assumed that the external perturbation $A_p(t)$ is turned on only for times $t\ge 0$.
	The terms involving $A_q(t_i)$ in the right-hand side describe how the external perturbation drives a response in the disk.
	The $B_q(t_i)$ in the inner square bracket represents the self gravity of the response.
	We omit these from the sum when comparing to test particle integration.
	
	\begin{figure*}
		\includegraphics[width=\textwidth]{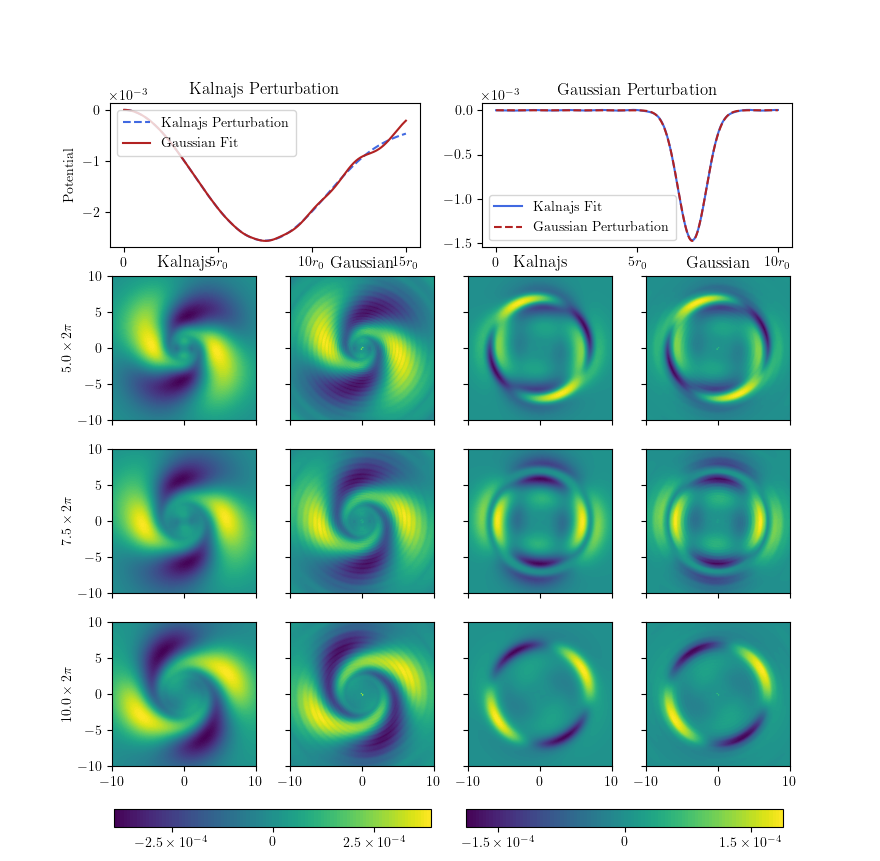}
		\caption{Comparing the effect of the choice of basis functions
			on the computed density response.  The top row shows two
			different $\ell=2$ potential perturbations applied at $t=0$
			together with their representations in the Kalnajs and Gaussian bases. The rows underneath plot the subsequent density evolution
			computed using the Kalnajs (left) and Gaussian (right)
			bases. There is strong agreement between the different basis functions, however imprints from the basis functions can be seen in the response - in particular the rings from the Gaussian basis functions.}
		\label{Fig: Basis Density Comparison}
	\end{figure*}			
	\subsection{Testing the effects of basis choice}
	\label{SubSec: Comparison of Different Basis Functions}
	In practice we must restrict the number of basis functions used to represent the disturbing potential and the subsequent response of the galaxy. Therefore our model does not in general see the `true' perturbing potential, nor can it predict all details of the response: instead, both are projected onto our finite basis. To test for systematic bias introduced by this projection one can compare predictions using different basis. \par 
	\autoref{Fig: Basis Density Comparison} shows the density response when the same perturbation is represented in two different basis functions. To ensure a fair comparison, both basis functions are expanded to the same order, namely $N_{max}=48$. The top row shows two different impulsive potentials $\psi\stim(\vx)$ that were applied to the disc at $t=0$, both with $\ell=2$. The perturbation on the left-hand side of the figure is spatially extended, that on the right-hand side more localised.
	The rows underneath each perturbation plot shows the time evolution of the corresponding
	density response, calculated using either the Kalnajs basis (left-hand column) or the
	Gaussian basis (right-hand column).\par 
	For the extended perturbation (left-hand panels), the Gaussian
	basis functions fail to reproduce the structure at large radii. This is because the central radius of the outer most Gaussian basis function is at $R=15R_{0}$ and therefore the Gaussian basis functions will fail to reproduce structure beyond this.  Due to the low density at larger radius, this effect of this on the subsequent evolution is minimal, as illustrated by the strong agreement between the Kalnajs and Gaussian density response. One can also see the imprint of the Gaussian basis functions, with the artificial rings imprinted on the density response. The fact that there is such strong agreement between two different basis functions implies that the basis functions expansion has been curtailed at an appropriate order.  \par	

	\begin{figure*}
		\includegraphics[width=\textwidth]{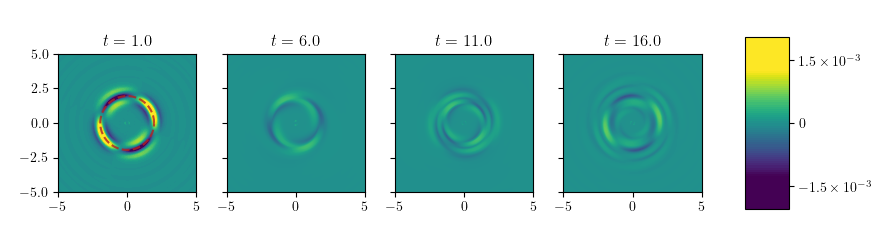}
		\caption{The quadrupole harmonic of the Green's function $\mathcal G_\rho(\vx,\vx',t)$ for a perturbation at $\vx'=(2r_0,0)$, calculated using the Gaussian basis. As this point mass can be decomposed into angular harmonics, this plot shows the quadrapole response of the disk. The red circle indicates the radius over which the perturbation is smeared when the quadrapole harmonic is taken, $r=2$.}
		\label{Fig: Green's Function}
		
	\end{figure*}
	\section{Green's Function}
	\label{Sec: Greens Function}

	From \autoref{Equ: Linear Volterra Equation of the Second Kind} it is a straight forward exercise to calculate the Green's function of the density response, $\mathcal{G}_{\rho}$. By definition this $\mathcal G_\rho(\vx,\vx',t-t')$ is the (density) response at location $\vx$ and time $t$ to an density perturbation $\rho\stim(\vx,t)=\delta(\vx-\vx')\delta(t-t')$.  By linearity then it follows that 
	\begin{equation}
	\rho\resp(\vx, t) = \int\d\vx'\d t'\, \mathcal{G}_{\rho}(\mathbf{x}, \mathbf{x}';  t-t')\rho^{e}(\mathbf{x}', t'),
	\label{Equ: Greens Function}
	\end{equation}
	\\ 
        which demonstrates that the Green's function is the fundamental object for solving initial value problems: one can use to to determine the evolution of the density and potential of the system subject to arbitrary perturbations.
	%

	We can use \autoref{Equ: Greens Function} to obtain $\mathcal{G}_{\rho}$ by setting $\rho\stim(\vx,\vt) = \delta(\mathbf{x}-\mathbf{x}) \delta(t-t)$ and then using the formalism developed in  \autoref{Sec: Dynamical Evolution} to evaluate the density evolution.  To do this we need to express the imposed Dirac delta perturbation as a linear combination of our basis functions.  Our basis set is truncated though, which means that the best we can do is to approximate
	\begin{equation}
	\delta(\vx-\vx') \delta(t-t') \approx \delta(t-t')\sum_{q=1}^{N_{max}}A_{q}\rho^{q},
        \label{eq:Diracapprox}
      \end{equation}
      as a linear combination of some number $N_{\rm max}$ of basis elements.
      With this approximation we have that
\begin{equation}
	\rho^{f}_{\mathbf{x}', t'}(\mathbf{x},t) = \int d\mathbf{x}'
	\mathcal{G}_{\rho} (\mathbf{x}, \mathbf{x}';t-t')\sum_{q=1}^{N_{nax}}A_{q}\rho^{q}(\textbf{x}')
\end{equation} 
and therefore
\begin{equation}
	\mathcal{G}_{\rho} (\mathbf{x}, \mathbf{x}';t-t') \simeq \rho^{f}_{\mathbf{x}', t'}(\mathbf{x},t). 
\end{equation}

We obtain the potential expansion coefficients~$A_q$ in~\eqref{eq:Diracapprox} by using~\eqref{eq:potproj} to fit a Keplerian $-1/|\vx-\vx'|$ potential with our Gaussian basis functions.  The corresponding density distribution is -- to with the resolution limit of our basis -- then the Dirac delta that appears in~\eqref{eq:Diracapprox}.
\autoref{Fig: Green's Function} plots the the quadrupole ($\ell=2$) harmonic of the resulting Green's function.
\section{Torque Acting on a Bar}
\label{Sec: Perturbations by an External Galaxy}

\begin{figure*}
	\includegraphics[width=\textwidth]{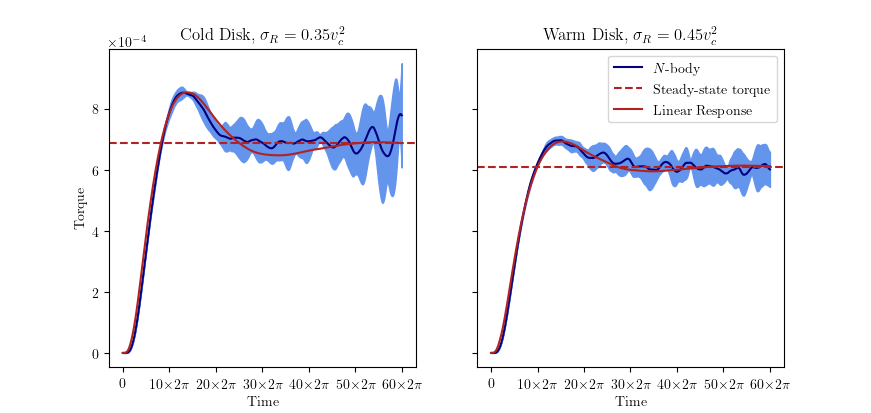}
	\includegraphics[width=\textwidth]{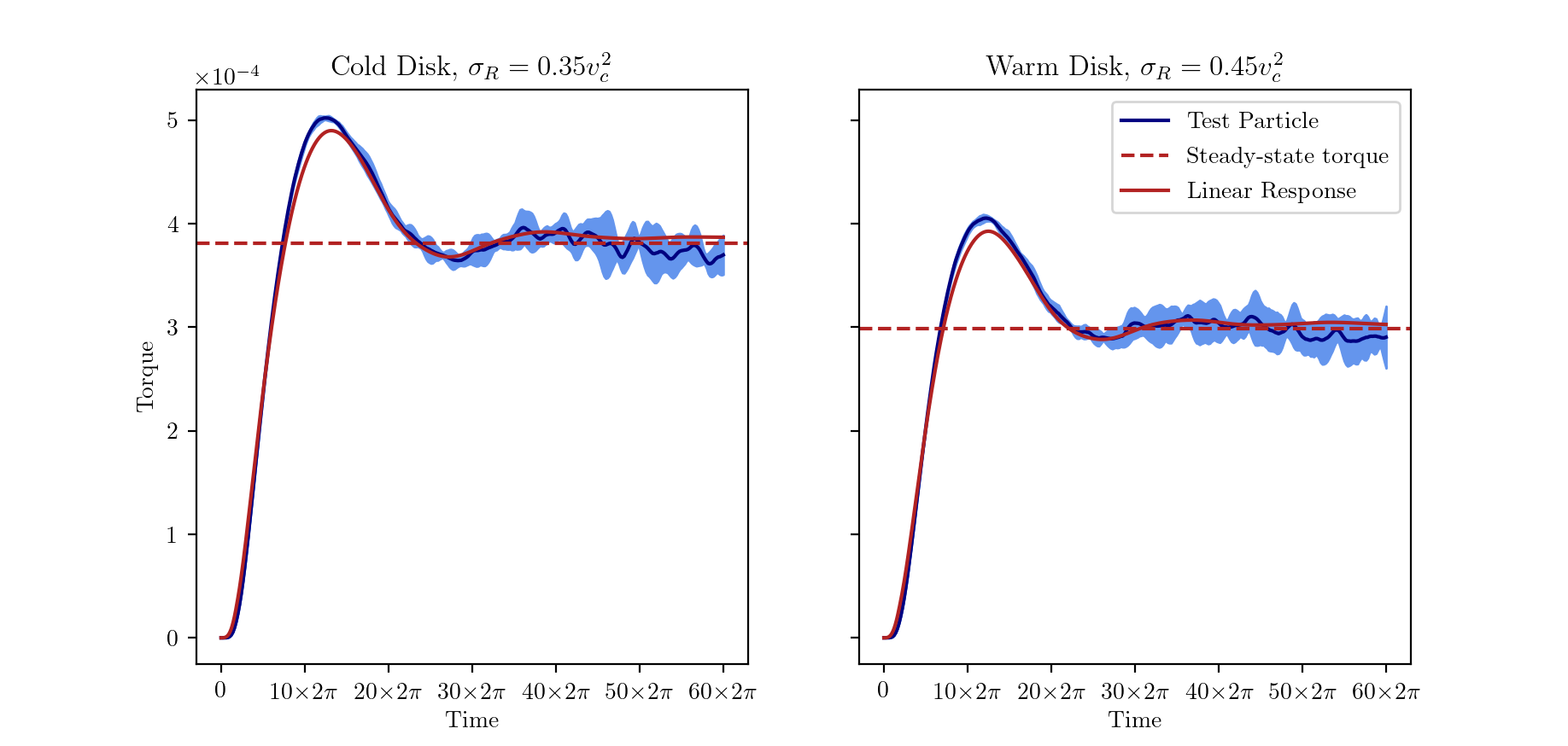}
	\caption{The frictional torque~\eqref{Equ: Bar torque} (red curve) exerted by discs of different temperatures when the bar~\eqref{eq:Bar1Ap} is imposed with pattern speed $\Omegap=1$.  The top row plots results when the self-gravity of the disc's response is included, while the bottom row shows corresponding results without self gravity for comparison: note the difference in scales.  The black curve shows torques measured from five $N$-body realisations, the blue shaded area indicating the standard deviation of these measurements.
          The horizontal dashed line is the steady-state torque~\eqref{Equ: New LBK}.}
	\label{Fig: Torque Different Temp}
	
\end{figure*}

We now use the formalism developed in \autoref{Sec: Dynamical Evolution} to study dynamical friction within the disk: we perturb the disc by a rotating bar-like potential and calculate the frictional torque that the disc's response exerts on the imposed bar.
Variants of this problem have been already studied in great detail, making use of both linear response theory and $N$-body simulations \cite[e.g.][,among others,]{tremaine1984dynamical, banik2021self, chiba2021oscillating, sellwood2006bar}. \par 
	\cite{tremaine1984dynamical} focused on friction in the secular adiabatic limit and showed that only trapped orbits contribute to the total torque on the bar in this regime. In doing so they generalised the work of \cite{lynden1972generating} and investigated the accuracy of the LBK formula. \cite{banik2021self} applied a similar formalism while relaxing the adiabatic assumption and found that the torque is a combination of two different parts: a transient component and a steady-state component. The transients are damped, not due to a dissipative process, but rather due to the phase mixing of different orbits. As the transients are inherently a time dependent phenomena, we can use the formalism laid out in this paper to study more closely the torque they apply to the bar. \par 
	For our bar model we take the lowest order mode of the Kalnajs basis functions (Appendix~\ref{App: Kalnajs Basis Functions}) rotating at constant pattern speed, $\Omegap=1$, i.e. 
	\begin{equation}
	A_{p}(t) =\begin{cases}
	\displaystyle \epsilon \massbar(t)\e^{-\i\Omegap t},& p = 0,
	\\
	\displaystyle 0,& \text{otherwise,}
      \end{cases}
      \label{eq:Bar1Ap}
	\end{equation}
	where $\massbar(t)$ is function that increases from 0 to~1 over time, and $\epsilon = 0.01$ sets the mass scale of the bar. \par 
	The disk's response will apply a torque to the bar, $\tau_{\rm b} =-\int d\vr \rho^{s}\left( \vr\times\partial\psi\resp/\partial\vr\right)$. 
          Using the expansion coefficients defined in \autoref{eq:ABexpansion}, this torque is given by
	\begin{equation}
	\tau_{\rm b}(\mathbf{x}_{\rm b}, t) = -i\ell \massbar\mathcal{E}^{pq}A_{p}(t)B^{*}_{q}(t).
	\label{Equ: Bar torque}
	\end{equation}
        For the bar growth we choose
	\begin{equation}
	\massbar(t) =\begin{cases}
	\displaystyle \sin^{2}\left(\frac{t}{4}\right),& t < 2\pi 
	\\
	\displaystyle 1, &\text{otherwise,}
	\end{cases}
	\end{equation}
        so that the bar grows over a single dynamical time at the position of the inner taper.
        We chose this short growth timescale to excite a strong transient response in the disk.
        
        \subsection{Results}

        For the linear calculation we use the Kalnajs basis functions, with $48$ terms in the basis expansion. We calculate the kernel for 1200 steps with a time step of $0.25$.  The resulting torque~\eqref{Equ: Bar torque} is plotted in \autoref{Fig: Torque Different Temp} for two different disc temperatures, $\sigma_R/v_{\rm c}=0.35$ and $0.45$ (left and right columns) and with (top row) and without (bottom row) the self-gravity of the response.

        As expected self-gravity is more important for cooler discs; with less random motion, stars find it easier to `unionise' and work together creating a coherent response, and hence a large torque.
        \autoref{Fig: Torque Different Temp} also shows that the torque from the time-dependent linear response is a combination of two parts: a transient response that damps out and a longer-term, constant contribution from the steady-state density wake that trails behind the bar.   These two contributions have already been found for  bars in halos \citep{banik2021self} and for the drag acting on a satellite pulled through the periodic cube \citep{magorrian2021stellar}.

 	\begin{figure*}
z		\includegraphics[width=\columnwidth]{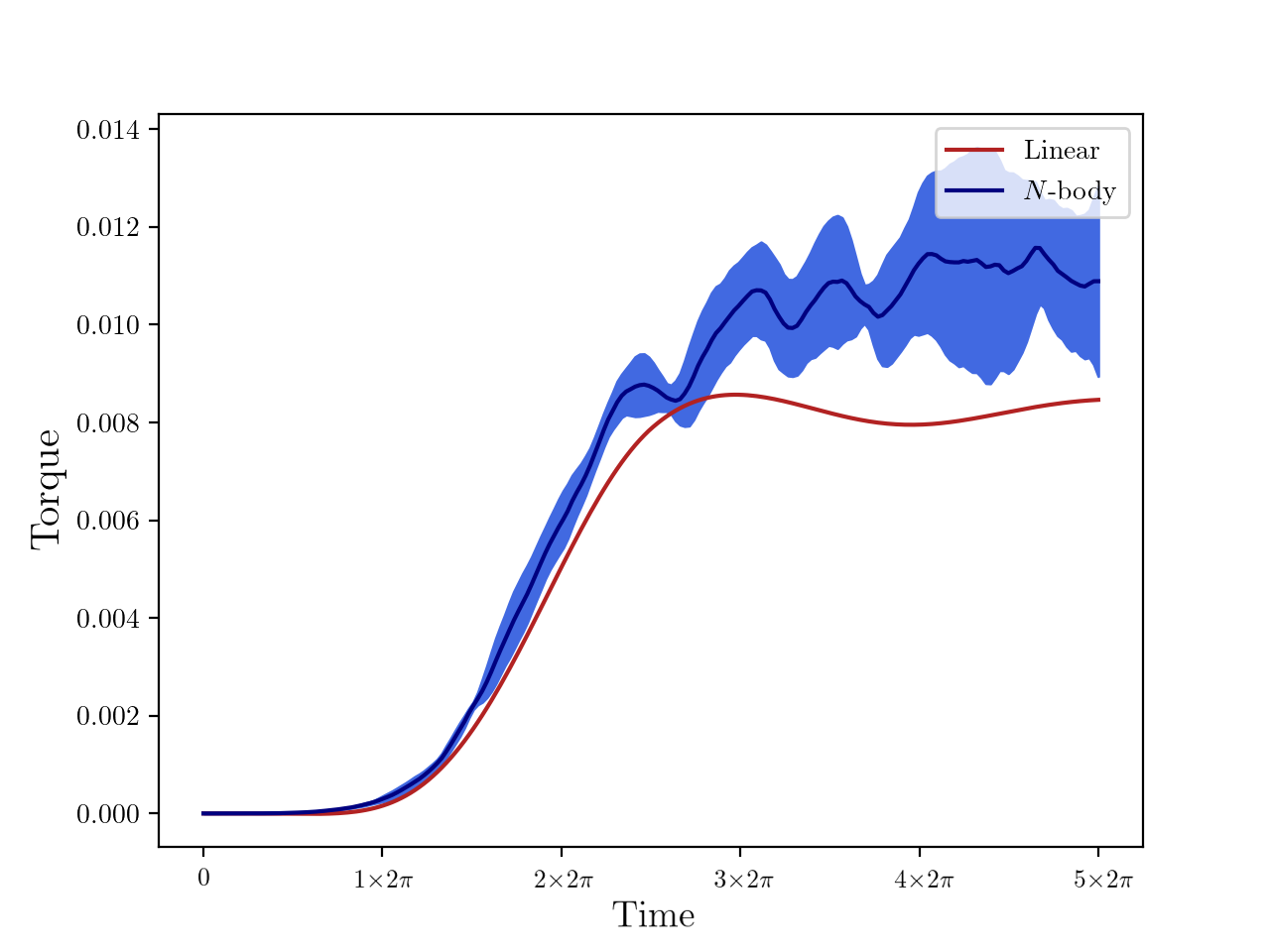}
		\includegraphics[width=\columnwidth]{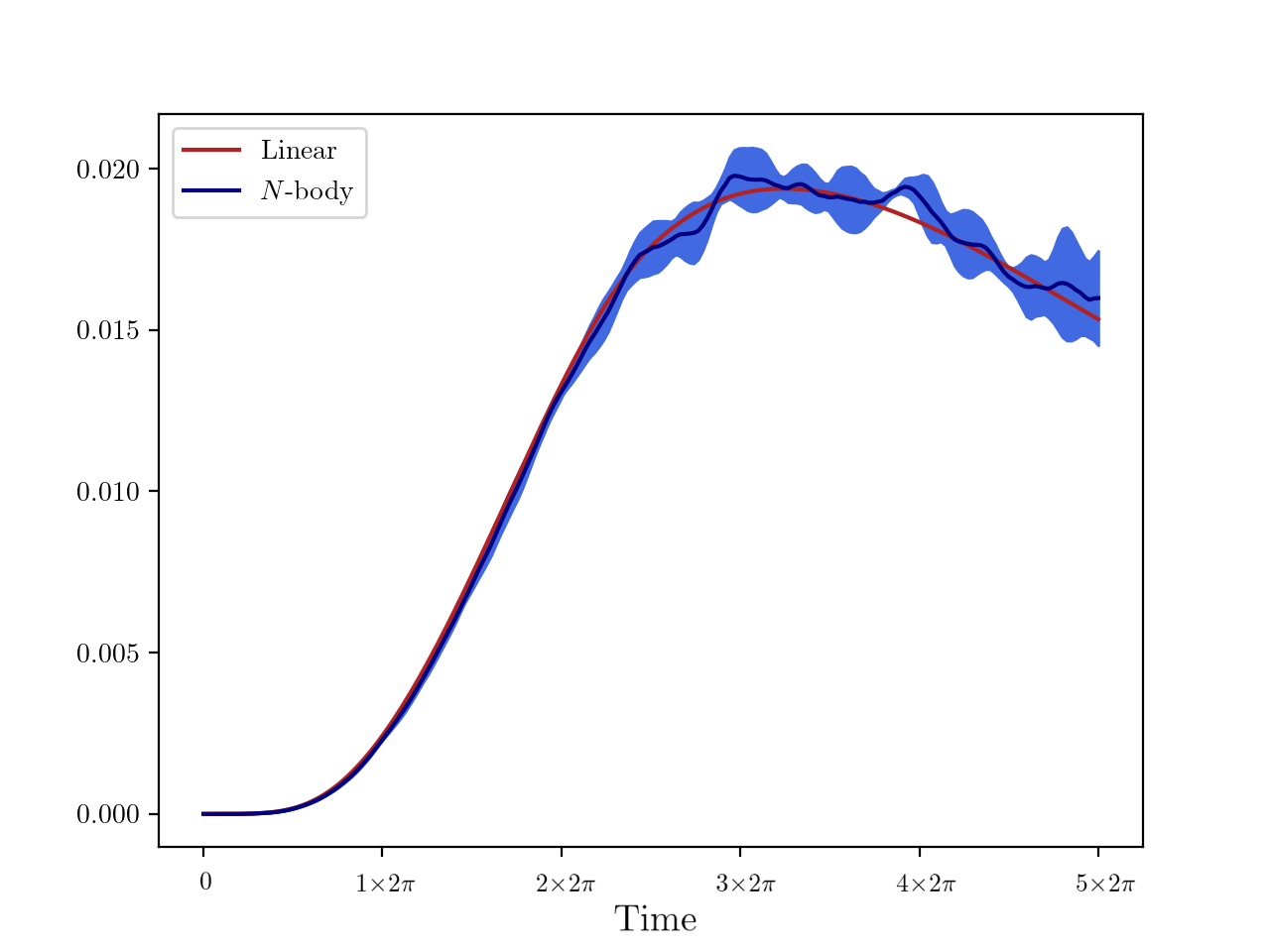}

		\caption{Evolution of the torque on our dumbell bar model (\autoref{Equ: Localised Bar Model}). The left-hand plot shows the torque on a fast dumbbell with $\Omegap=0.5t_0^{-1}$, the right-hand plot $\Omegap=0.1t_0^{-1}$. The disagreement between $N$-body results and our linearized calculation in the fast dumbbell arises because a significant fraction of the disc's matter moves on resonantly trapped orbits, which are not accounted for in the linear theory.}
		\label{Fig: Bar torque N body}
	\end{figure*}

        \subsection{An expression for the steady-state torque}
        \label{SubSec: An expression for the steady-state torque}
        We can apply the method used by  \cite{magorrian2021stellar} to calculate this longer term, steady-state torque.  After the bar has developed the stirring potential is simply
	\begin{equation}
	A_{p}(t) = \hat{A}_{p}\e^{-\i\Omegap t},
	\label{Equ: Perturbation Assumption}
	\end{equation} 
        in which the coefficients $\hat A_{p}$ are constant.  We assume that the disc is stable, so that, after the transients have died away and the dynamical equilibrium established, its response will be
	\begin{equation}
	B_{p}(t) = \hat{B}_{p}\e^{-i\Omegap t},
	\label{Equ: Response Assumption}
	\end{equation}   
        where the coefficients $\hat B_{p}$ are independent of time and dependent linearly on the imposed $\hat A_{p}$.
	Making the change of variable $\tau = t - t'$ in \autoref{Equ: Linear Volterra Equation of the Second Kind} gives 
	\begin{equation}
	\mathcal{E}^{pq}\hat{B}_{q} = \mathcal{M}^{pq}(\Omegap)\left[\hat{A}_{q} + \hat{B}_{q}\right].
		\label{Equ: Modal Bar}
	\end{equation}
	where $\mathcal{M}^{pq}(\Omegap)$ is the response matrix (\autoref{Equ: Response Matrix}). It is unsurprising that the response matrix should appear, as by making the assumptions in \autoref{Equ: Perturbation Assumption} and \ref{Equ: Response Assumption} we have carried out a modal analysis. \par 
	We can now combine our expression for the response in \autoref{Equ: Modal Bar} with our general expression for the torque on the bar in \autoref{Equ: Bar torque} to obtain,
	\begin{equation}
	\tau_{\rm b}(t\rightarrow\infty) = -i\ell\mathcal{E}^{pq}\hat{A}_{p}\left[(\mathcal{E}-\mathcal{M})^{-1}_{qr}\mathcal{M}^{rs}\hat{A}_{s}\right]^{*}.
	\label{Equ: New LBK}
	\end{equation}
        an expression that has already been obtained by \cite[][equation 53]{WeinbergSelfgravitatingresponsespherical1989} taking a slightly different route.
As the method outlined in \autoref{Sec: Dynamical Evolution} gives the evolution of  potential-density coefficients (rather than the perturbation of the DF) , we naturally express the torque in these coefficients, leading to a concise expression for the torque. 
This torque is second order in the size of the perturbation, $m_{b}$, due to the product of the $\hat{A}_{p}$ terms \citep[e.g.][]{tremaine1984dynamical, banik2021self}.
Unlike those authors, however, but like \cite{WeinbergSelfgravitatingresponsespherical1989}, the expression~\eqref{Equ: New LBK} accounts for the self-gravity of the disc's response via the resonant denominator.

	In order to calculate the response matrix~\eqref{Equ: Response Matrix} we follow the approach taken by \cite{fouvry2015secular}. As most of the individual component of the response matrix are needed in order to calculate the evolution kernel (\autoref{Equ: Kernel}), obtaining the response matrix is not too challenging. 	The horizontal dashed lines in Figure~\ref{Fig: Torque Different Temp} plot the torque obtained from \autoref{Equ: New LBK}, showing that it agrees very well with the fully time-dependent linear response calculations.  We stress, however, that the real strength of the method outlined in this paper is to calculate the latter.\par
	\subsection{Comparison with $N$-body simulation}
	The results we have just presented are based on a number of approximations. For example, we had to introduce a truncated basis expansion to expand the density and potential response and we neglet the second-order $[f,\psi]$ term in the CBE.  In order to test these assumptions, we compare our linear response calculation to $N$-body simulations, an independent test of our method. Appendix \ref{App: Details of Test-Particle and N-Body Integrators} gives details of how we set up and run our $N$-body simulations.
        For each panel of Figure~\ref{Fig: Torque Different Temp} we have run five $N=5\times10^5$ realisations of the disc's response and plot the mean and standard deviation of the torque measured from these simulations.  The strong agreement between the two different methods justifies the assumptions that we have made in our linear response calculations.  \par 

	Although the linearity assumption works well for the problem of calculating the net torque on the bar in this particular case, there are nevertheless underlying nonlinear effects that have not been accounted for, most notably orbit trapping.  These can be particularly important for systems with constant, or almost-constant, pattern speeds.  We now modify our bar potential to highlight these.
        
	\subsection{Effect of Resonant Trapping}
	\label{SubSec: Effect of Resonant Trapping}
	
	\begin{figure}
		\includegraphics[width=.5\textwidth]{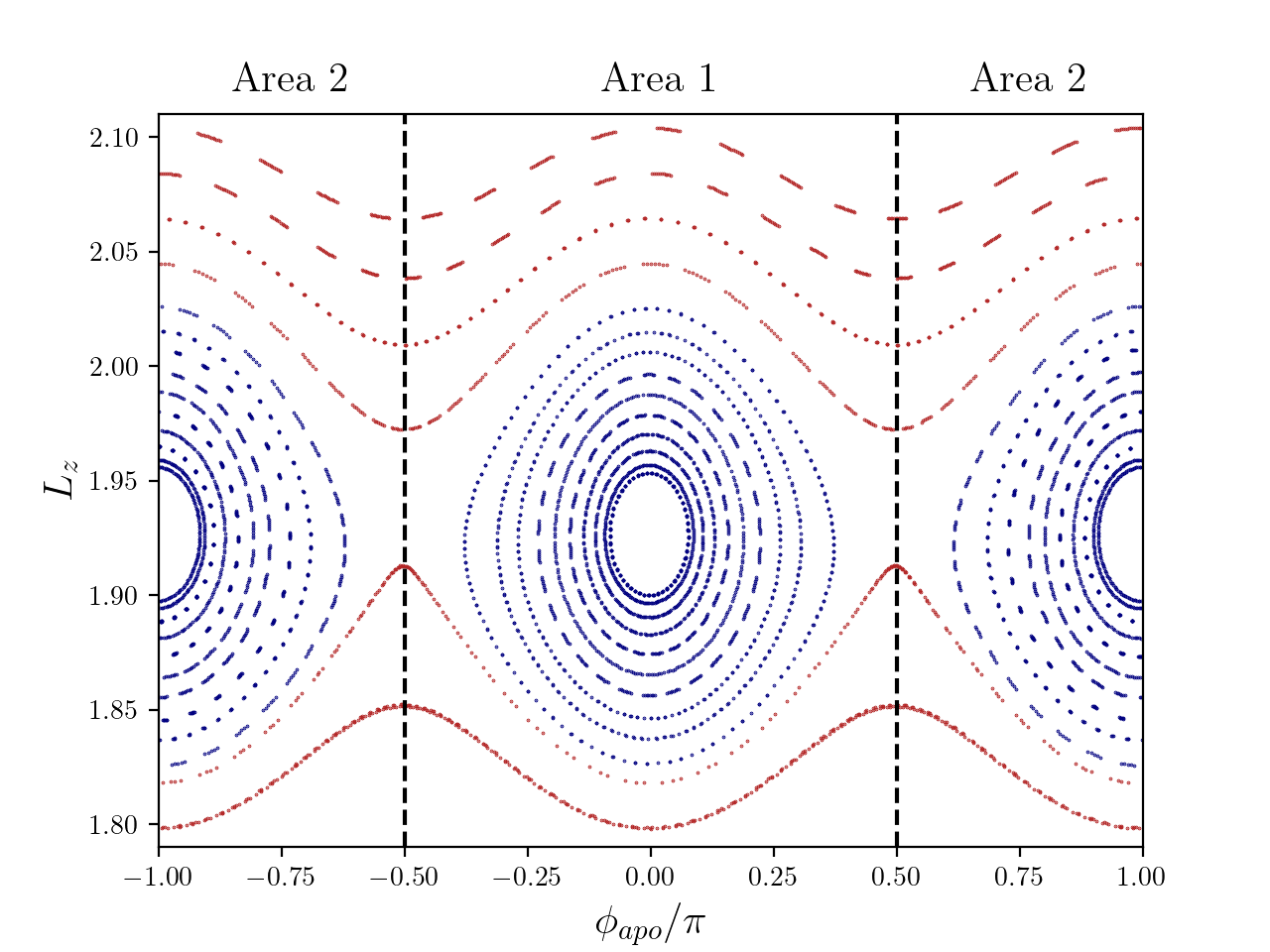}
		\caption{Surface of section for test particles in a disc perturbed by the rapidly rotating dumbbell potential~\eqref{Equ: Localised Bar Model} with corotation radius is at $2R_0$ ($\Omegap=0.5t_0^{-1}$).  The surface of section is constructed by plotting the angular momentum~$L_z$ versus the angle relative to the bar, $\phi_{apo}$,  each time any orbit passes through apocentre. The trapped orbits, plotted in red, are radically altered by the perturbation, causing the assumptions of linear theory to breakdown. All orbits are sampled such that they have a constant Jacobi integral $H_{\rm J}=0.23v_{\rm c}^2$, which corresponds to a circular orbit at $R=2.1R_0$.} 
		\label{Fig: Orbit Sections}
	\end{figure}
 Linear perturbation theory computes the response by integrating perturbations along unperturbed orbits. Resonant trapping will drastically change the orbits, breaking this assumption. In the case where we expect resonant trapping to be important, we expect our linear theory to fail.\par 
 To make the effect of orbital trapping as stark as possible we use a new, dumbbell bar model where we consider two localised blobs of matter orbiting opposite each other,
	\begin{equation}
	\rho_{\rm p}(R, \phi, t=0) =\epsilon\frac{ m_{\rm p}(t)}{2\pi \Delta_{ p}^{2}}\e^{-\left(R-R_{\rm p}\right)^{2}/(2\Delta_{\rm p}^{2})}\cos\left(2\phi\right),
	\label{Equ: Localised Bar Model}
	\end{equation}
	where we take $R_{\rm p} = 2R_{0}$ as the blobs' radii and $\Delta_{\rm p} = 0.1R_{0}$ are their characteristic widths. 
	%
        The extended nature of the bar model used in the previous section means that many particles are perturbed, of which only a small fraction are resonantly trapped. For the more localised perturbation of~\eqref{Equ: Localised Bar Model}, however, there is a very strong response from particles at radii $R\sim R_{\rm p}$ that corotate with the dumbbell.  If the pattern speed is chosen such that its co-rotation resonance is located at $r_{\rm p}$ --  i.e., if we choose $\Omegap \simeq v_{c}/r_{\rm p}$ -- then the largest possible fraction of perturbed particles will be trapped, leading to a large discrepancy between $N$-body simulations, that include resonant orbits, and our linear theory, which does not. \par  

	To illustrate the importance of this effect, we plot the torque acting on the disc for two different pattern speeds in \autoref{Fig: Bar torque N body}. For a dumbbell with $\Omegap=0.1t_0^{-1}$ the corotation radius is at $10R_0$, well outside the ``reach'' of the dumbbell.  Very few orbits are trapped and there is strong agreement between the linearized calculation and the intrinsically non-linear $N$-body simulations.  On the other hand, spinning the dumbbell up to $\Omegap=0.5t_{0}^{-1}$ puts corotation at $2 R_{0}$, the approximate density peak of the tapered Mestel disc, and large number of trapped orbits means that the linear calculation disagrees with the $N$-body results.

	To illustrate the existence of trapped orbits, we plot surfaces of sections of test particle orbits in the potential of the rotating dumbbell.  We introduce a co-rotating $(x',y')$ frame with the blobs located at $(x',y')=(\pm 2R_{\rm p},0)$.   The new coordinate system is a rotation of the old coordinate system, 
	\begin{equation}
	\begin{split}
	&x' = x \cos\left(\Omega_{m}t\right) + y\sin\left(\Omega_{m}t\right),\\
	&y' = y\cos\left(\Omega_{m}t\right) - x\sin\left(\Omega_{m}t\right), 
	\end{split}
	\end{equation}
	and the generalised momenta in the co-rotating frame then take the form,
	\begin{equation}
	\mathbf{p}_{\mathbf{x'}} = \dot{\mathbf{x}} +\Omega_{m}\hat{\mathbf{z}}\times \mathbf{x}. 
	\end{equation}
	In this new coordinate system the Jacobi integral, $H_{\rm J} = H - \bm\Omega_{m}\cdot\mathbf{L}$, becomes a conserved quantity. \par  
        \autoref{Fig: Orbit Sections} shows a surface of section 
      for $H_{\rm J} =  0.23v_{c}^{2}$ (corresponding to a circular orbit at $R=2.1$) in the  fast rotating dumbbell ($\Omegap=0.5t_0^{-1}$).  To create the surface of section we have run a number of orbits having this $H_{\rm J}$ for 250 radial periods, plotting values of angular momentum~$L_z$ versus azimuthal angle $\phi_{\rm apo}$ at apocentre passages.  It clearly shows the existence of orbits trapped around $\phi_{\rm apo}=0$ and $\phi_{\rm apo}=\pm\pi$.
      
 %
 We can use the form of \autoref{Fig: Orbit Sections} to construct a quick estimate of the total fraction of particles that are trapped.  We split $\phi_{apo}$ into two areas: Area $1$ ($-\pi/2 \leq\phi_{apo} \leq \pi/2$) and Area $2$ ($\phi_{apo}<-\pi/2$ or $\phi_{apo} > \pi/2$). Trapped (librating) orbits will remain in either area, whereas streaming (circulating) orbits will pass between the two. As our $N$-body code samples from the tapered Mestel disc, to calculate the fraction of orbits trapped by the perturbation, all we need to do is count the number of particles that remain constrained to either one of the two areas during a run. Doing this for the simulations presented in Figure~\ref{Fig: Bar torque N body}, we find no trapped orbits, but approximately $10\%$ of orbits trapped in the fast dumbbell ones. As the fraction of trapped orbits is comparable to the discrepancy between linear theory and $N$-body, we conclude that the disagreement is driven by trapping. Trapped orbits increase the effective mass of the bar, leading to a larger perturbation to the disc and hence a larger torque. 

	In future work it will be important to contend with resonant orbits. There are different ways to do this, but the most elegant approach is to transform into slow and fast angles and to solve the resulting pendulum equation \citep[e.g.][]{tremaine1984dynamical, chiba2021oscillating}. \par 
	\section{Conclusion}
	\label{Sec: Conclusion}

        In this paper we have presented a practical scheme for following the response of 2D stellar discs to internal and external perturbations by following \cite{julian1966non,Seguin1994,PichonAubert2006} and \cite{murali1999transmission} among others in recasting the linearised CBE as a Volterra integral equation of the first kind, which is easy to solve numerically.
        \cite{rozier2022constraining} have recently presented a similar application to galactic halos.

        There are many studies that ``solve'' the integral equation by Fourier transforming in time and considering the response in frequency~$\omega$ instead.
        There are situations in which this is the natural approach, such as, e.g., testing for stability, or when the galaxy is subject to a periodic perturbation.  Nevertheless in most situations the direct approach has a couple of advantages over these frequency-based methods:
        \begin{enumerate}
        \item it avoids the difficulties in dealing with the denominator of the response matrix~\eqref{Equ: Response Matrix};
        \item it becomes straightforward to include the effect of the galaxy's response on an external perturber, making it possible to follow, e.g., the decay of a satellite's orbit directly (see, e.g., \cite{magorrian2021stellar} for a simple example).
        \end{enumerate}
        Moreover, we demonstrate in Section~\ref{SubSec: An expression for the steady-state torque} that our direct machinery can still be used to obtain the frequency-space response, essentially by postponing the Fourier transform in time to the end of the calculation.

        We have applied the machinery to calculate the dynamical friction torque on an imposed bar.  Introducing the bar excites modes that damp quickly, leaving a state in which a steady density wake is established trailing the bar and exerting an constant torque.  The colder the disc, the stronger the disc's density response and the stronger the resulting torque.  Turning off the self gravity of the disc's response reduces the torque by approximately a factor of two.
        
        We find good agreement with the torques measured from $N$-body simulations, as long as the perturbation does not capture many stars on resonantly trapped orbits.  The main advantage of the linear response calculation over $N$-body simulation is the resolution with which it enables us to study the response: for example, in Figures~\ref{Fig: Torque Different Temp} and~\ref{Fig: Bar torque N body} the $N$-body torques are much noisier than the linear-response ones.  Moreover, the linear-response calculation is faster: once the kernels (which depend on the background DF, the basis and the choice of timestep) have been precomputed then solving equation~\eqref{Equ: Linear Volterra Equation of the Second Kind} is a matter of seconds, allowing the effect of different perturbations to be studied with ease.  For the calculations of Section~\ref{Sec: Perturbations by an External Galaxy} the time taken to compute the kernels was comparable to the time needed to run a single $N$-body realisation.

        $N$-body simulation has the advantages, however, of being a much better understood technique and of naturally capturing nonlinear effects (provided, of course, that $N$ is high enough).  The success of the linearized calculation depends on the completeness of the choice of potential--density pairs \citep{kalnajs1976dynamics} used to represent the response.  We have considered two quite different sets of these, finding that they produce comparable results.  An open question is how best to choose them for a given problem (e.g., to minimise the number of pairs needed to reproduce the self-gravitating response correctly).
        No matter how cleverly they are chosen though, they cannot remedy the biggest blindspot of the linearized formalism: the effects of orbit trapping by resonances (Section~\ref{SubSec: Effect of Resonant Trapping}).  Although there are many applications that are unaffected by this blindspot, an interesting challenge would be to adapt some variant of the machinery presented here that accounts for trapped orbits.

	\bibliographystyle{mnras}
	\bibliography{Linear_Stability_Bib} 
	\appendix
	
	\section{Details of test-particle and $N$-body integrators}
	\label{App: Details of Test-Particle and N-Body Integrators}
	$N$-body simulations provide an informative test of the linear assumption made in \autoref{Sec: Dynamical Evolution}. In this appendix we outline the implementation of our $N$-body code, before comparing it to our linear response calculations. The strong agreement between the two fundamentally different calculations suggests the successful implementation of \autoref{Equ: Linear Volterra Equation of the Second Kind}.  \par 
	\subsection{Sampling the DF}
	Our first task is to sample from the DF described in \autoref{Sec: Application to razor-thin axisymmetric discs}, we do this in two parts: initially, we sample the positions of the particles, before sampling their generalised momenta. For each particle we must sample four coordinates, $(R_{i}, \phi_{i}, v_{R,i}, v_{\phi,i})$ and assign a weight $\mu$. We sample in polar coordinates due to the angular symmetry of the DF. For velocity sampling we work in $(v_{R}, v_{\phi})$ coordinates, rather than $(E,L)$ as the former, along with the generalised coordinates $(R, \phi)$, forms a canonical coordinate system, and therefore does not require the inclusion of a Jacobian factor.  \par 
	Initially, we sample $(R_{i}, \phi_{i})$. We start by calculating the cumulative radial distribution function,
	\begin{equation}
	D(<R) = \int_{0}^{R}(2\pi R)dR \int d\mathbf{v}F(R, v_{R}, v_{\phi}),
	\end{equation}
	where the distribution function, $F$, is given in \autoref{Equ: Distribution}. Once this is calculated we can sample $r_{i}$ using the inverse transform sampling method. Due to the angular symmetry, $\phi_{i}$ is trivially sampled, $\phi \sim \mathcal{U}(0, 2\pi)$. \par 
	We now turn our attention to the velocity sampling, the radial velocity is easily sampled, $v_{R} \sim \mathcal{N}(0, \sigma_{R}^{2})$, however the angular velocity is harder to sample, due to the non-trivial tapers. \par
	To sample $v_{\phi,i}$, we use a rejection sampling method. At this point in the sampling algorithm we have three out of four phase space coordinates, namely $R_{i}$, $\phi_{i}$ and $v_{R,i}$, this reduces the DF to one dimension, $F(v_{\phi}| R_{i}, \phi_{i}, v_{R,i})$. For rejection sampling, we assume that this constrained DF is bounded $v_{\phi} \in [0, v_{\phi, \text{max}}]$ and $F(v_{\phi})\in[0,\mathcal{M}]$. The upper cut off on $v_{\phi}$ is chosen as $v_{\phi,max} = v_{c}+10\sigma_{R}$. We then pick two uniformly distributed values $v_{\phi,i}$ and $y_{i}$, distributed according to $v_{\phi}\sim\mathcal{U}(0, v_{\phi,max})$ and $y \sim\mathcal{U}(0, \mathcal{M})$. We accept $v_{\phi,i}$ if,
	\begin{equation}
	\label{Equ: Rejection sampling}
	\begin{split}
	y_{i} \leq F(v_{\phi,i}| R_{i}, \phi_{i}, v_{R,i}) &\implies \text{$v_{\phi,i}$ is accepted,} \\
	y_{i} > F(v_{\phi,i}| R_{i}, \phi_{i}, v_{R,i}) &\implies \text{$v_{\phi,i}$ is rejected.} 
	\end{split}			
	\end{equation}
	We continue to resample until the criteria in \autoref{Equ: Rejection sampling} is met. This algorithm is made more efficient by picking $\mathcal{M}$ as close to the maximum of $F(v_{\phi})$ as possible. As both $0  \leq T_{\text{inner}}, T_{\text{outer}} \leq 1$, we choose $\mathcal{M}$ as the maximum value of the untapered Mestel DF (\autoref{Equ: Mestel DF}) given the values already sampled.  \par
	Once we have sampled the four components for each particle, we transform into Cartesian coordinates - the coordinate system in which our $N$-body simulation will work - using the relations,
	\begin{subequations}
		\begin{equation}
		x_{i} = R_{i}\cos\phi_{i},
		\end{equation}
		\begin{equation}
		y_{i} = R_{i}\sin\phi_{i},
		\end{equation}
		\begin{equation}
		v_{x,i} = v_{R,i}\cos(\phi_{i}) - v_{\phi,i}\sin(\phi_{i}),
		\end{equation}
		\begin{equation}
		v_{y,i} = v_{R,i}\sin(\phi_{i}) + v_{\phi,i}\cos(\phi_{i}).
		\end{equation}
	\end{subequations}
	\par
	To each particle we assign a mass $\mu = M_{\text{disc}}/N$, where $M_{\text{disc}}$ is the mass of the active part of the disc and $N$ are the number of particles in the simulation. \par 
	\subsection{$N$-Body Evolution}
	Once we have sampled the particles for the $N$-body simulation, we must generate the potential they source, and which they move in. To do this we split the potential into three parts: 
	\begin{enumerate}
		\item An axisymmetric contribution $\Phi_{0}(R)$ that is sourced by the background Mestel disc and is calculated analytically using $\Phi_{0}(R) = v_{c}^{2}\log(R/R_{0})$.
		\item The potential sourced by the fluctuations in the disc, $\psi^{f}(R,\phi)$.
		\item The external perturbation that will `stir' the disc, $\psi^{e}(R, \phi)$. 
	\end{enumerate}
	To calculate $\psi^{f}$ we assign the particles to a square mesh of size $N_{\text{grid}}\times N_{\text{grid}}$, with uniform grid spacing $\Delta R$, using a cloud-in-cell interpolation scheme. This density is then filtered in order to isolate the Fourier harmonic we are looking for, i.e. if we are interested in the $\ell=1$ response, we calculate the $\ell =1$ density perturbation. This filtered density is then turned into a potential using the `doubling up' method described in section 2 of \cite{binney2011galactic}. A slight complication arises when calculating the $\ell = 0$ response as as after the filtering step two parts remain, the $\ell =0$ perturbation due to the density fluctuations and the potential due to the background disc, i.e. a combination of (i) and (ii). This background will be subject to fluctuations that will drive dynamics, and therefore we want to work with the analytic background of a Mestel disc. To subtract this component out, we calculate the potential caused by the particles when they are initially sampled and subtract this from filtered potential, leaving the potential due the density fluctuation. To this we then add the analytic background when doing the leapfrog integrator.  \par 
	To generate the external perturbation, we initially calculate the coordinates of the perturbation in the choose basis, i.e. we calculate $A_{p}(t)$ using \autoref{Equ: Potential-density pair def}. These coefficients can then be used to generate a grid representing the external perturbation (and can also be fed into \autoref{Equ: Linear Volterra Equation of the Second Kind} to calculate the density response via the linear response method set out in \autoref{Sec: Dynamical Evolution}). \par 
	For test particle simulations, we turn off the self-consistent response, in effect setting the active fraction to zero. \par
	Once we have the total acceleration, from the background potential, the self-consistent response and the external perturbation, we use a first order leapfrog integrator to evolve the particles in time. \par
	In order to compare the predictions made by \autoref{Equ: Linear Volterra Equation of the Second Kind} and $N$-body we must calculate the expansion coefficients of the density fluctuations, $B_{p}(t)$, from the $N$-body simulation. To do this we run the code twice with the same initial conditions: initially, the `background' run, in which the external perturbation is turned off and secondly, the `foreground' run, including the external perturbation. For each foreground and background simulation we construct the foreground and background distribution function, $f_{f}$ and $f_{b}$ respectively,
	\begin{equation}
	f_{\text{f/b}} = \sum_{n = 1}^{N} \mu\delta^{2}(\mathbf{x}-\mathbf{x}_{i, \text{f/b}})\delta^{2}(\mathbf{v}-\mathbf{v}_{i, \text{f/b}}),
	\end{equation}
	where $\mu = M_{\text{disc}}/N$ is the weight of each particle and $N$ is the total number of particles, ensuring that the mass of the disc is the same in the $N$-body and linear response calculation. The density fluctuation can then be calculated from the marginalised Klimontovich DFs,
	\begin{equation}
	\delta \rho(\mathbf{x}) = \int d^{2}\mathbf{v}\left[f_{\text{f}}(\mathbf{x}, \mathbf{v}) - f_{\text{b}}(\mathbf{x}, \mathbf{v})\right].
	\end{equation}
	Combining the density fluctuation along with the bi-orthogonality relation (\ref{Equ: Potential-density pair def}), we can calculate  $B_{p}(t)$,
	
	\begin{equation}
	B_{p}(t) = -\mathcal{E}_{pq}^{-1}\int d^{2}\mathbf{x}\left[\psi^{q}(\mathbf{x})\right]^{*}\delta\rho(\mathbf{x}). 
	\end{equation}
	\par
	For the $N$-body simulation we set $R_{0} = {v_{c}} = G = 1$, as they are scale parameters. For all results presented here we use $R_{max} = 20R_{0}$, with a uniform time step $\Delta t = 10^{-3}R_{0}/v_{c}$. For our background DF we set $\sigma = 0.35$, $\nu_{t} = 4$, $\mu_{t} = 5$, $R_{i} = 1$, $R_{o} = 10$ and $\xi = 1/2$. We use $N = 10^{5}$ particles. For the potential solver we take a maximum radius of $26r_{0}$ with a box size of $N_{\text{grid}} = 120$. When taking the Fourier modes of the density we use $120$ radial rings, each with $720$ steps. The softening length is taken as $0.18 R_{0}$. 	\par 
	\section{Kalnajs Basis Functions}
	\label{App: Kalnajs Basis Functions}
	\begin{figure}
		
		\includegraphics[width = 0.5\textwidth]{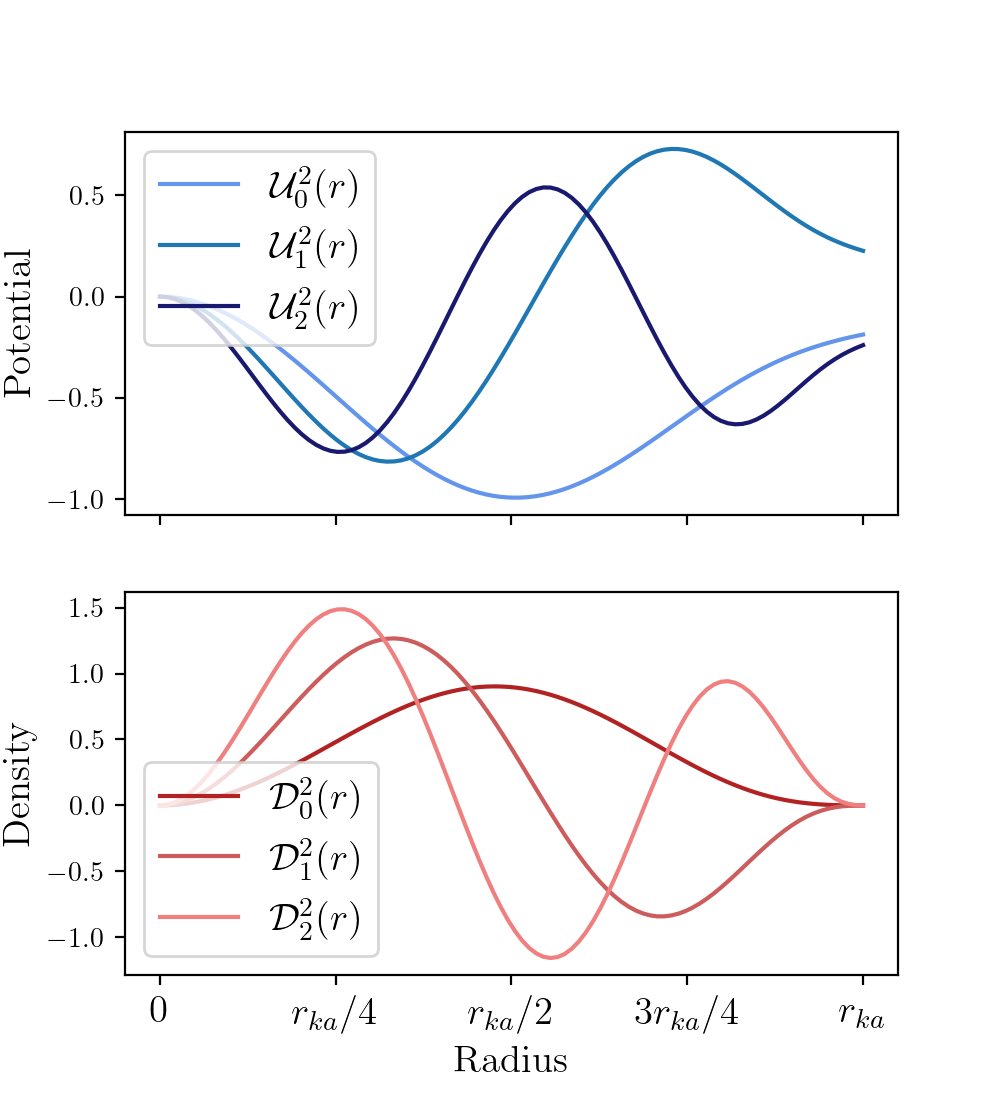}
		\centering
		\caption{The radial part of the Kalnajs basis functions. The top row shows the potential functions, $\mathcal{U}^{\ell}_{n}(R)$, and the bottom row the density functions, $\mathcal{D}^{\ell}_{n}(R)$. Each row plots the first three $n$ modes for a $\ell = 2$. The radial scale length is given by $R_{ka}$ which we take $R_{ka}=20$ for this work.} 
		\label{Fig: Kalnajs Basis Functions}
	\end{figure}

	In the following section we use the notation of \cite{fouvry2015secular}. The radial basis function, first introduced in \autoref{Sec: Dynamical Evolution}, have two indices: a radial index, $n_{p}\geq0$ and an angular index, $\ell_{p}\in \mathbb{Z}$.  The `Kalnajs basis' first described in \cite{kalnajs1976dynamics} have two more parameters: $k_{\textit{Ka}} \in \mathbb{N}$, which sets the highest polynomial power and a scale radius $r_{\textit{Ka}} \in \mathbb{R}^{+}$. For this work we kept $k_{\textit{Ka}} = 4$ and $R_{\textit{Ka}} = 20$. \par  
	The radial part of the basis functions assumes that $\ell \geq 0$, therefore wherever $\ell$ is used below it is presumed to be $|\ell|$. In the following we express $r$ as a dimensionless quantity $ R = R/R_{\textit{Ka}}$. A limited number of the basis functions are plotted in \autoref{Fig: Kalnajs Basis Functions}. The potential, $\mathcal{U}^{\ell}_{n}(R)$, and density, $\mathcal{D}^{\ell}_{n}(R)$, radial functions are
	\begin{subequations}
		\begin{equation}
		\begin{split}
		\mathcal{U}^{\ell}_{n}(R) = &-\sqrt{\frac{G}{R_{\textit{Ka}}}}\mathcal{P}(k_{\textit{Ka}}, \ell, n) R^{\ell} \times \\
		&\sum_{i=0}^{k}
		\sum_{j=0}^{n}\alpha_{\textit{Ka}}(k_{\textit{Ka}}, \ell, n,i,j)R^{2i+2j}.
		\end{split}
		\end{equation}
		\begin{equation}
		\begin{split}
		\mathcal{D}^{\ell}_{n}(R) = & \frac{(-1)^{n}}{\sqrt{G R_{\textit{Ka}}^{3}}}\mathcal{S}(k_{\textit{Ka}}, \ell, n)(1-R^{2})^{k_{\textit{Ka}}-1/2}  
		R^{\ell} \times \\
		&\sum_{j=0}^{n}\beta_{\textit{Ka}}(k_{\textit{Ka}},\ell, n, j)(1-R^{2})^{j}, 
		\end{split}
		\end{equation}
	\end{subequations}
	Using the Pochammer symbol, $[a]_{i}$ (defined below), the normalisation factors for the potential are 
	\begin{subequations}
		\begin{equation}
		\begin{split}
		\mathcal{P}(k,\ell,n) = &\left[\frac{[2k +\ell +2n+1/2]\Gamma[2k+\ell+n+1/2]}{\Gamma[\ell+n+1/2]\Gamma^{2}[\ell+1]\Gamma[n+1]}\right]^{1/2} \times \\
		& \Gamma[\ell+n+1/2]^{1/2}, 
		\end{split}
		\end{equation}
		\begin{equation}
		\begin{split}
		\alpha_{\textit{Ka}}(k,\ell, n,i,j) = 
		& \left[\frac{[-k]_{i}[\ell+1/2]_{i}[2k+\ell+n+1/2]_{j}}{[\ell+1]_{i}[1]_{i}[\ell+i+1]_{j}[\ell+1/2]_{j}[1]_{j}}\right] \times \\
		&[i+\ell+1/2]_{j}[-n]_{j}.
		\end{split}
		\end{equation}
		\label{Equ: Global normalisation}
	\end{subequations}
	The normalisation for the density are,
	\begin{subequations}
		\begin{equation}
		\begin{split}
		&\mathcal{S}(k,\ell, n) = \frac{\Gamma[k+1]}{\pi \Gamma[2k+1]\Gamma[k+1/2]} \times \\
		& \left[\frac{[2k+\ell +2n +1/2]\Gamma[2k+n+1]\Gamma[2k+\ell+n+1/2]}{\Gamma[\ell+n+1/2]\Gamma[n+1]}\right]^{1/2}, 
		\end{split}
		\end{equation}
		\begin{equation}
		\beta_{\textit{Ka}}(k,\ell,n, j) = \frac{[2k+\ell+n+1/2]_{j}[k+1]_{j}[-n]_{j}}{[2k+1]_{j}[k+1/2]_{j}[1]_{j}.}
		\end{equation}
		\label{Equ: Term normalisation}
	\end{subequations}
	The Pochammer symbol is defined as,
	\begin{equation}
	[a]_{i}=\begin{cases}
	1, & \text{if $i=0$},\\
	a(a+1)...(a+n-1), & \text{if $i > 0$}. 
	\end{cases}
	\end{equation}
	It is worth noting that the calculation of the normalisation coefficients is very costly. \par 
	In general when calculating kernels with the Kalnajs basis functions we take the following parameters. The number of terms in our basis expansion $N_{max} = 10$. When calculating the action-angle representation of the potential-density pairs, $\mathcal{W}^{m{1}}_{\ell m_{2} n}(\mathbf{J})$, we do so on a uniform grid in $(R_{+},R_{-})$ coordinates, with step size $\Delta R_{grid}=0.15R_{0}$ and $R_{+},r_{-}\leq R_{ka}$. To obtain the kernel from these we sum over $m_{1}$, therefore we calculate the action-angle representation for of the basis functions for $-7\leq m_{1} \leq7$. We check for stability in all of these parameters and find them to best balance computational speed and accuracy. However, in \autoref{SubSec: Comparison of Different Basis Functions}, when calculating kernels to compare with Gaussian basis functions we increase the number of basis functions so that $N_{max} = 48$. Due to the finer structure of higher order modes we require a finer grid of $\mathcal{W}$, so we take $\Delta r_{grid}=0.08R_{0}$. All the other parameters remain the same.    
	\section{Gaussian Basis Functions}
	\label{Sec: Gaussian Basis Functions}
	\begin{figure}
		\includegraphics[width = .5\textwidth]{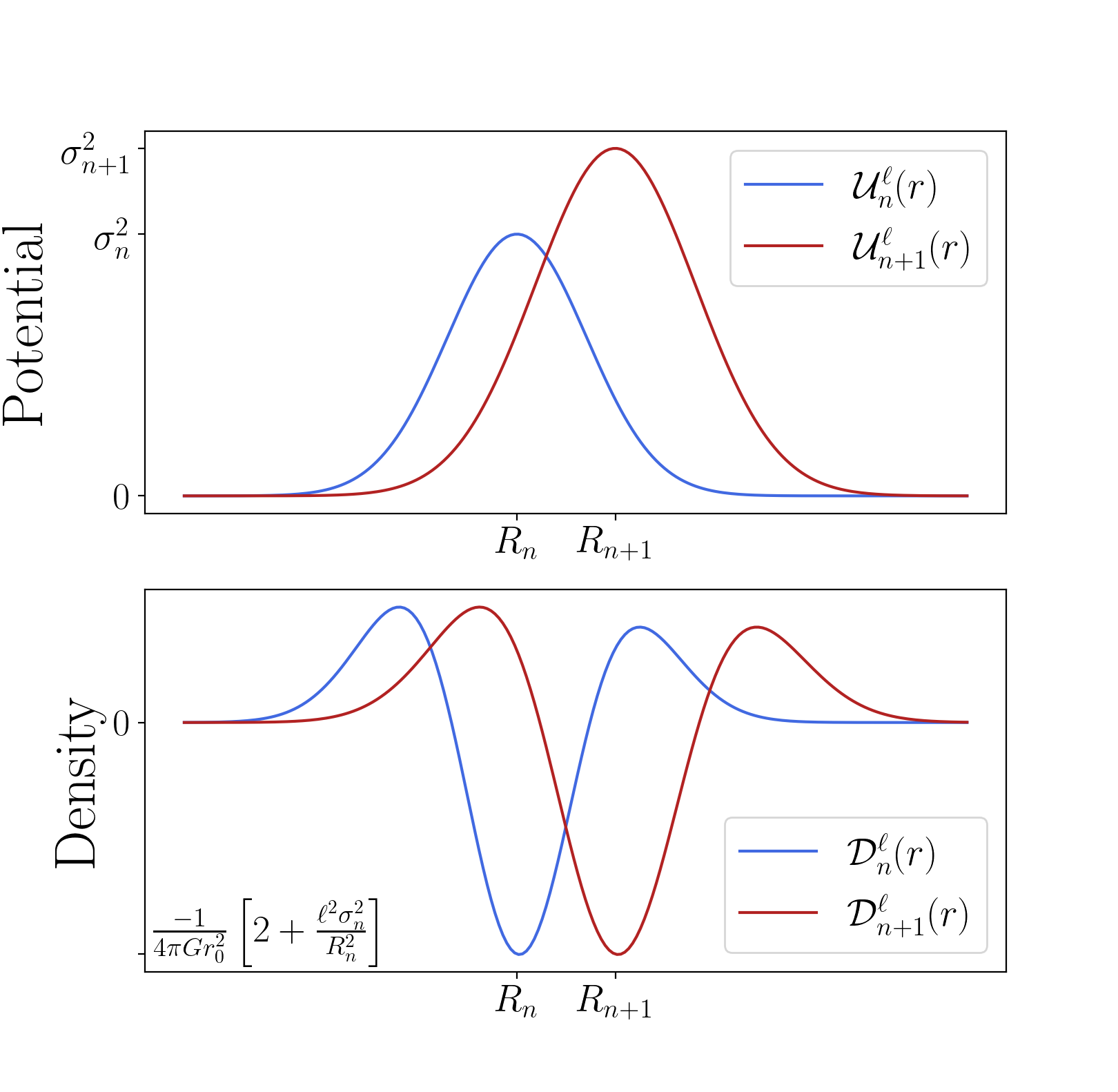}
		\centering
		\caption{Plot of two successive Gaussian basis functions. We logarithmically space the peaks of the basis functions and choose the widths, $\sigma_{n}$, such that $\sigma_{n} = R_{n+1} - R_{n}$.}
		\label{Fig: Spiral basis functions}
	\end{figure}
	%
	%
	To calculate the action representation of the basis functions, we follow the procedure laid out in \autoref{Sec: Application to razor-thin axisymmetric discs}. We plot two successive Gaussian basis function in \autoref{Fig: Spiral basis functions}.\par 
	For the Gaussian basis function we use 25 basis functions with an inner basis function centred at $R_{0} = 0.15$ and an outer basis function centred at $R_{48} = 15$. We choose these parameters to ensure that we can resolve the inner taper of our disc. To expand the Gaussian basis functions in potential density pairs, we use again use $-7\leq m_{1} \leq7$. However due to the localised nature of inner basis function we decrease our grid spacing of $\mathcal{W}$, so that $\Delta r_{grid}=0.08$. Again we check for stability when all of these parameters are varied. 

\end{document}